# Study of n-γ discrimination in low energy range (above 40 keVee) by charge comparison method with a BC501A liquid scintillation detector

CHEN Yong-Hao(陈永浩)[1,2]　CHEN Xi-Meng(陈熙萌)[1]　ZHANG Xiao-Dong(张小东)[1]
LEI Jia-Rong(雷家荣)[2]　AN Li(安力)[2;1)]　SHAO Jian-Xiong(邵剑雄)[1]　ZHENG Pu(郑普)[2]
WANG Xin-Hua(王新华)[2]　ZHU Chuan-Xin(朱传新)[2]　HE Tie(何铁)[2]　YANG Jian(阳剑)[2]

[1] School of Nuclear Science and Technology, Lanzhou University, Lanzhou 730000, China
[2] Institute of Nuclear Physics and Chemistry, Chinese Academy of Engineering Physics, Mianyang 621900, China

**Abstract:** A VME-based experiment system for n-γ discrimination using the charge comparison method was established. A data acquisition program for controlling the programmable modules and processing data online via VME64X bus was developed through the use of LabVIEW. The two-dimensional (2D) scatter plots of the charge in the slow component vs. the total charge of recorded pulses from $^{241}$Am-Be and $^{252}$Cf neutron sources were presented. The 2D scatter plots of the energy vs. the ratio of the charge in the slow component to the total charge of the pulses were presented at the meantime. The quality of n-γ discrimination was checked by the figure-of-merit, and the results showed good performance of n-γ discrimination at low energy range. Neutrons and γ-rays were separated above 50 keVee (electron-equivalent energy). The quality of n-γ discrimination have been improved compared with others' results at 5 energies (150, 250, 350, 450, 550 keVee).

## 1 Introduction

Neutron detection is of great importance in many basic research studies and applications, such as nuclear reactor control, nuclear radiation protection, and nuclear structure research. It has been found that all neutron fields coexist with associated γ-rays background, arising as a result of reactions of the neutrons with materials in the environment and as direct byproducts of the primary reaction producing the neutron field. A challenge to detect neutrons is the associated γ-rays background, thus discriminating neutrons against γ-rays background plays an essential role. The use of BC501A liquid scintillation detectors gives an efficient way to discriminate neutrons against and γ-rays by means of the pulse shape discrimination (PSD) [1-3].

Charge comparison method is one of effective PSD methods to perform the n-γ discrimination [4-6]. It is usually done by comparing the charge integration of the current pulse over two different time intervals using charge-to-digital converter (QDC). The charge comparison method can be technically or electronically implemented in variety of ways. For example, A. Lavagno et al [7] and J. Cerny et al [8] used the common technique of comparing the fractional charge in the tail with the total charge integrated by QDC. A. Jhingan et al [9] used two methods to perform the charge comparison. The first method was the common method as above. In second method they replaced total charge integration by the shaped dynode pulse, which was fed to a peak-sensing ADC, and then compared it with the integrated fractional charge of the anode pulse. M. Nakhostin [10] and K.A.A. Gamage et al [11] used a digital oscilloscope to directly digitize the signals from the anode of the photomultiplier (PMT), then used an algorithm to process the sampled signals offline.

Most of these previous methods were based on the NIM or CAMAC modules, or digitizers, while an experiment system based on the programmable VME modules would be presented in this paper.

In this paper, a VME-based experiment system for n-γ discrimination by the charge comparison method was established. A LabVIEW program for controlling the programmable modules via the VME bus and processing data on-line were developed. The system was tested with $^{241}$Am-Be and $^{252}$Cf neutron sources. The results showed excellent performance of the n-γ discrimination, the energy threshold of n-γ discrimination could go down to 50 keVee. The figure-of-merits (FOMs) at 5 energies were calculated and compared with other group's results, the quality of n-γ discrimination was comparatively improved.

## 2 Experimental details

In this work, a cylindrical BC501A liquid scintillator (3″ in diameter and 2″ in height) coupled to a PMT (9265KB of ET Enterprises) with silicon oil was used to detect neutrons and γ-rays. In order to inhibit the scattering background of neutrons and γ-rays, the experiment was arranged in a spacious experimental hall (26.3 m in length, 11.4 m in width and 14 m in height), the neutron sources ($^{241}$Am-Be source with the intensity of $2.5 \times 10^6$ n/s, $^{252}$Cf source with the intensity of $1.6 \times 10^6$ n/s) and detector were put at the center of it. The detector was supported by a thin steel bracket and positioned perpendicularly to the ground, the distance between the front surface of the scintillator and ground is 3.8 m. The neutron source was suspended 90 cm away from the front surface of the scintillator on its central axis. This kind of arrangement could minimize the scattering neutron background.

The block diagram of the electronic circuit and the corresponding time relation of the signals are shown in Fig. 1. The CAEN V812 is a 1-unit wide VME module housing 16 constant fraction discriminator (CFD) channels. The CAEN V792N is a 1-unit wide VME module housing 16 QDC channels with 12-bit resolution. The CFD, QDC, High Voltage(HV) are all controlled and adjusted by a PC via the VME bus through the LabVIEW program.

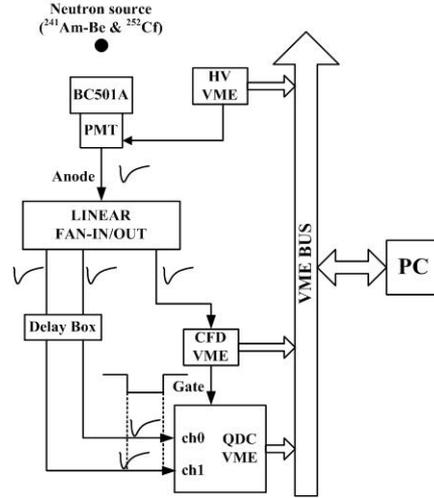

Fig. 1. The block diagram of the electronic circuit employed in the present work (LINEAR FAN-IN/OUT-PHILLIPS 740; Delay Box- ORTEC DB463; HV-CAEN V6533; CFD-CAEN V812; QDC-CAEN V792N).

As Fig. 1 shows, each signal coming from the anode of PMT was split into 3 similar signals by linear fan-in/fan-out. One of them was sent into CFD to generate a Gate signal for QDC, the other 2 signals were delayed and sent into a multievent QDC for integrating. Adjusting the delay so that one signal was totally integrated to $Q_{total}$, the other was partly integrated to $Q_{slow}$ (only the falling edge of the signal was integrated). In the case of Fig. 1, the signal fed to the ch0 was integrated to $Q_{total}$, the signal fed to the ch1 was integrated to $Q_{slow}$. It was found that the n-γ discrimination performed best when the output width of CFD was set to 148 ns, the delays of the two signals were set to 54 ns and 14 ns for $Q_{total}$ and $Q_{slow}$ respectively.

## 3 Energy calibration

The energy calibration for the system must be done before measuring the neutron sources. Organic scintillator, because their constituent elements have low atomic numbers, have very low photoelectric interaction probabilities. They produce pulses when exposed to γ-rays, but are almost never used for γ-ray spectroscopy. The interactions that occur are primarily single or multiple Compton scatterings that can deposit only a fraction of the incident γ-ray energy, so full-energy peaks are not observed in typical height spectra [12] (Fig 2).

As the light output of electron is known to be linear with its energy in the range of 0.04MeV≤Ee≤1.6 MeV [13], the Compton electrons induced by γ-rays are used to perform the energy calibration. For this reason, the energy is usually described in keVee or MeVee, where ee stands for electron-equivalent energy unit. The maximum energy of Compton electron, $E_{emax}$, could be calculated as [14]

$$E_{e\max} = E_\gamma \left( \frac{\frac{2E_\gamma}{m_0 c^2}}{1 + \frac{2E_\gamma}{m_0 c^2}} \right), \quad (1)$$

where $E_\gamma$ is the energy of the incident γ-rays, $m_0 c^2$ is the rest-mass energy of the electron (0.511 MeV).

A $^{133}$Ba γ-ray source with the intensity of $3.86 \times 10^4$ Bq was used to calibrate the system. The energy of its γ-rays is 0.356 MeV, according to the equation (1) the corresponding maximum energy of the Compton recoil electron is 0.207 MeV. The Compton recoil electron spectrum of $^{133}$Ba is shown in Fig. 2.

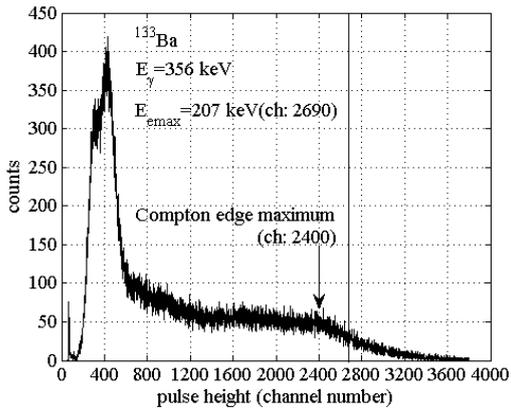

Fig. 2. The Compton recoil electron spectrum of $^{133}$Ba used for performing the energy calibration.

The energy calibration is done using the channel number at 75% of the Compton edge maximum [2]. It is around 2690 channel in the case of Fig. 2.

## 4 Results and discussion

The two-dimensional (2D) scatter plots of the n-γ discrimination with $^{241}$Am-Be and $^{252}$Cf neutron sources are shown in Fig. 3 and Fig. 4.

Fig. 3 shows the 2D n-γ discrimination spectra of $^{241}$Am-Be neutron source at low energy range. The Fig. 3(a) and 3(b) are based on the same experimental data actually, but shown in different ways. The Fig. 3(a) represents the 2D plot of $(Q_{slow}/Q_{total}) \times 100$ vs. energy, while the Fig. 3(b) represents the 2D plot of $Q_{slow}$ vs. $Q_{total}$.

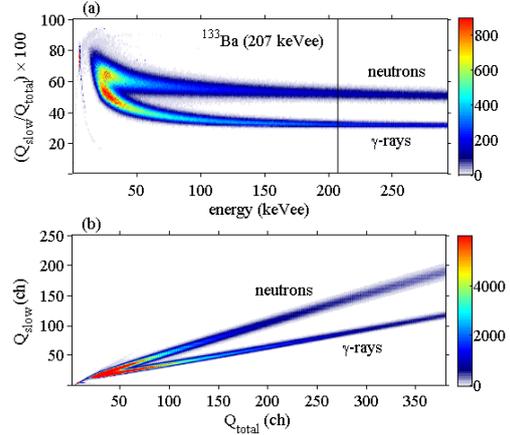

Fig. 3. (color online) 2D scatter plots of n-γ discrimination with $^{241}$Am-Be neutron source.

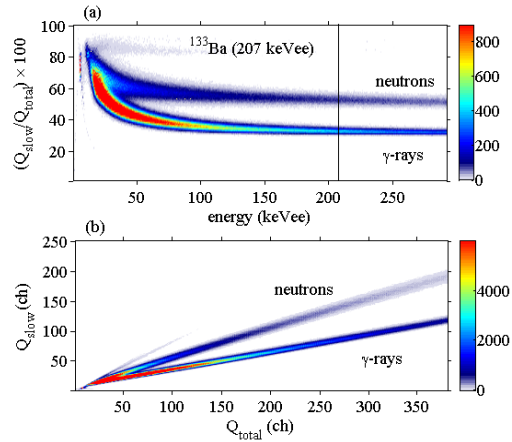

Fig. 4. (color online) 2D scatter plots of n-γ discrimination with $^{252}$Cf neutron source.

Fig. 4 shows the 2D n-γ discrimination spectra of $^{252}$Cf neutron source at low energy range, it is shown in the same way as Fig. 3. It can be seen from Fig. 3(a) and Fig. 4(a) that the neutron and γ-ray events are almost completely separated above 50 keVee.

As Fig. 2 shows, the total channel number of QDC's output is 3800. As a consequence, the size of Fig. 3(a) and Fig. 4(a) is 3800×100. In the same way, the size of Fig. 3(b) and Fig. 4(b) should be 3800×2500. However, such big size is time-consuming for processing, so we re-bin the spectra by summing every ten channels into one channel, then the size of Fig. 3(b) and Fig. 4(b) is changed into 380×250.

The quality of n-γ separation would be checked by the FOM, which is defined as

$$FOM = \frac{S}{FWHM_n + FWHM_\gamma}, \quad (2)$$

Where S is the separation between the peaks of the neutron and γ-ray events. The $FWHM_n$ and $FWHM_\gamma$ are the full-width at half-maximum of neutrons and γ-rays peaks respectively [15].

The procedure of calculating the FOMs is exemplified with the spectrum of $^{241}$Am-Be source at 40 keVee (Fig. 5). A LabVIEW program is developed to do the three point smoothing firstly, then find out the values of S, $FWHM_n$, and $FWHM_\gamma$, finally the FOM is calculated according to the equation (1).

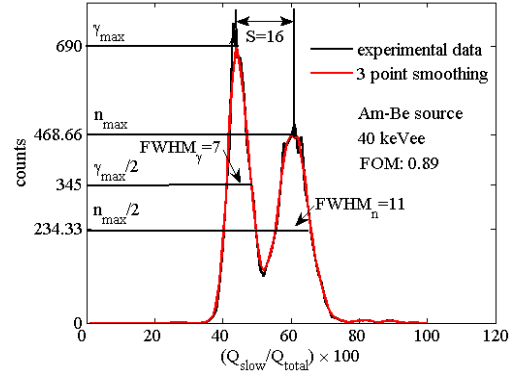

Fig. 5. (color online) The procedure of calculating the FOM of $^{241}$Am-Be source at 40 keVee.

In order to evaluate the lower energy threshold of n-γ discrimination by this system, the qualities of n-γ discrimination at 4 energies are shown with FOMs and neutron peak-to-valley ratios (Table 1). The FOM is used to describe the separation quality of neutrons and γ-rays. The neutron peak-to-valley ratio is defined by the ratio of neutron peak over the valley between the neutron and γ-ray peak.

Table 1 The quality of n-γ discrimination at 4 energies with $^{241}$Am-Be and $^{252}$Cf source

| Neutron source | Energy(keVee) | FOM | Neutron peak-to-valley ratio |
| --- | --- | --- | --- |
| $^{241}$Am-Be | 40 | 0.89 | 3.58 |
| | 50 | 1.13 | 8.69 |
| | 60 | 1.19 | 17.50 |
| | 70 | 1.35 | 28.13 |
| $^{252}$Cf | 40 | 0.74 | 2.17 |
| | 50 | 1.13 | 4.77 |
| | 60 | 1.20 | 12.44 |
| | 70 | 1.58 | 42.85 |

It can be seen through Table 1 that the neutrons and γ-rays are separated clearly above 50 keVee for both $^{241}$Am-Be and $^{252}$Cf neutron sources because the FOMs are above 1, the neutron peak-to-valley ratios are more than 5 ($^{241}$Am-Be source) or close to 5 ($^{252}$Cf source). It can be concluded that the energy threshold of n-γ discrimination is extended down to 50 keVee.

In order to compare the n-γ discrimination quality with others' results, the 2D n-γ discrimination spectrum of $^{241}$Am-Be neutron source in higher energy range (compared with Fig. 3(a)) was obtained (Fig. 6), the energy calibration was performed by a standard $^{60}$Co γ-ray source with the intensity of 3.62×10$^4$ Bq.

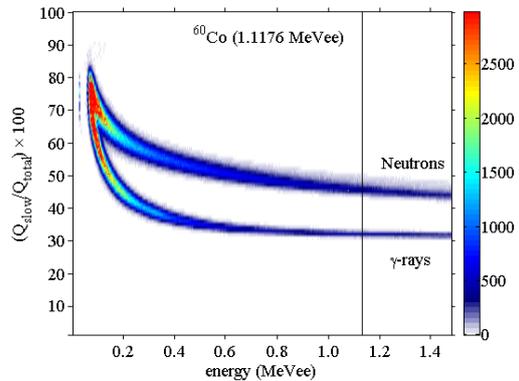

Fig. 6. (color online) 2D scatter plot of n-γ discrimination with $^{241}$Am-Be neutron source in higher energy range.

The FOMs at 5 energies (150, 250, 350, 450, 550 keVee) were calculated and compared with

M. Nakhostin's results [2], as shown in Fig. 7, which were achieved by the digital charge comparison method. The FOMs in this work were improved from 1.25 to 2.43 when the energy increased from 150 keVee to 550 keVee. The quality of n-γ discrimination is better than M. Nakhostin's results because the FOM is comparatively higher at each energy.

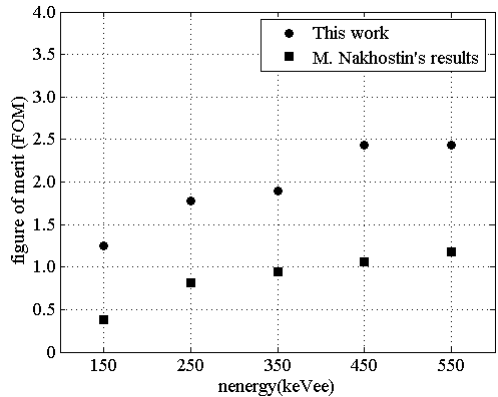

Fig. 7. Comparison of FOMs at 5 energies with M. Nakhostin's results.

The key points for improving the n-γ discrimination quality compared with M. Nakhostin's results are mainly based on the following: 1. The delays of the signals were adjusted to a optimal level so that the n-γ discrimination performed best. 2. The used liquid scintillator size of $\Phi3''\times2''$ was proved to be better than $\Phi2''\times2''$ for n-γ discrimination [16]. 3. The used QDC has a high resolution bit, 12-bit, compared with the 8-bit in the reference, so the results were more accurate and showed good performance of n-γ discrimination.

## 5 Conclusions

In this paper, an experiment system for n-γ discrimination by the charge comparison method based on programmable VME modules was presented. A program for controlling and reading out the programmable modules via VME64X bus as well as processing data online was developed by the use of LabVIEW. The system was tested with $^{241}$Am-Be and $^{252}$Cf neutron sources, and the results showed an excellent n-γ discrimination quality. The neutrons and γ-rays were separated clearly above 50 keVee in terms of FOM and neutron peak-to-valley ratio, which indicated that the energy threshold of n-γ discrimination was extended down to 50 keVee. The quality of n-γ discrimination was improved compared with M. Nakhostinl's results at 5 energies.

The modules used in the experiment house multichannel inputs (the CFD and QDC house 16 channel inputs, but only 1 and 2 channels are needed for one detector), so this system can reduce the complexity of the electronics when large numbers of neutron detectors are involved.